\documentclass[aps, prd, twocolumn, a4paper]{revtex4-1}
\usepackage{hyperref}
\usepackage{graphicx}
\usepackage{amsmath}
\usepackage{xcolor}
\usepackage{float}
\newcommand{\be}{\begin{equation}}
\newcommand{\ee}{\end{equation}}
\newcommand{\ba}{\begin{eqnarray}}
\newcommand{\ea}{\end{eqnarray}}
\newcommand{\nn}{\nonumber\\}

\begin{document}
\title{Dissociation of heavy quarkonia in an anisotropic hot QCD medium in a Quasi-Particle Model}
  \author{Mohammad Yousuf Jamal $^{a}$}
\email{mohammad.yousuf@iitgn.ac.in}
 \author{Indrani Nilima $^{b}$}
\email{nilima.ism@gmail.com}
 \author{Vinod Chandra $^{a}$}
\email{vchandra@iitgn.ac.in}
\author{Vineet Kumar Agotiya$^{b}$}
\email{agotiya81@gmail.com}
 \affiliation{$^a$Indian Institute of Technology Gandhinagar,  Gandhinagar-382355, Gujarat, India}
 \affiliation{$^b$Centre for Applied Physics, Central University of Jharkhand Ranchi, India, 835 205}
\begin{abstract}

The present article is the follow-up work of, Phys.\ Rev.\ D {\bf 94}, 094006 (2016), where we
have extended the study of quarkonia dissociation in (momentum) anisotropic hot QCD medium. As evident
by the experimentally observed collective
flow at RHIC and LHC, the momentum anisotropy is present at almost all the stages after the collision and therefore, it
is important to include its effects in the analysis. Employing the in-medium (corrected) potential while considering the anisotropy (both oblate and 
prolate cases) in the medium, the thermal widths and the binding energies of the heavy quarkonia states (s-wave charmonia
and s-wave bottomonia specifically, for radial quantum numbers n = 1 and 2) have been determined. 
The hot QCD medium effects have been included employing a quasi-particle description.  The presence of
anisotropy has modified the potential and then the thermal widths and the binding energies of these states in a significant manner.
The results show a quite visible shift in the values of dissociation temperatures as compared to the isotropic case. 
Further, the hot QCD medium interaction effects suppress the dissociation temperature as compared to the case where we consider the medium as
a non-interacting ultra-relativistic gas of quarks (anti-quarks) and gluons.
\end{abstract}
\maketitle

\section{Introduction}

 In the ultra-relativistic heavy-ion collision experiments at RHIC and LHC, it has been inferred that the 
 Quark-Gluon-Plasma (QGP) formed, functions more like a perfect fluid rather than a non-interacting ultra-relativistic
 gas of quarks (anti-quarks) and gluons ~\cite{expt_rhic, expt_lhc, Heinz:2004qz}. This is because of the fact that, the QGP
 possess a robust collective property that could be quantified in terms of the flow harmonics.
Among the other important signatures based on the experimental observations, quarkonia ($Q{\bar Q}$) suppression
  has also been suggested as a clear probe of the QGP formation in the collider experiments ~\cite{McLerran:1986zb, Back:2004je}.
  As observed in the experiments, it accentuates the plasma aspects of the medium, for example, Landau damping
  ~\cite{Landau:1984}, color screening~\cite{Chu:1988wh} and the energy loss~\cite{Koike:1991mf}.
  
After the discovery of $J/\psi$ ( a bound state of $c {\bar c}$), ~\cite{Aubert:1974js, Augustin:1974xw}, in 1974,
both the experimental as well as the theoretical studies of heavy quarkonia has become an interesting topic for the researchers to investigate.
A pioneering research, the dissociation of quarkonia due to the color screening in the deconfined medium with
finite temperature, was first carried out by  Matsui and Satz~\cite{Matsui:1986dk}. Thereafter, a large number of excellent articles
have been published that envisioned several essential refinements in the study of 
quarkonia~\cite{Mocsy:2004bv, Agotiya:2008ie, Patra:2009qy, Datta:2003ww, Brambilla:2008cx}. 
   
Quarkonia is the color singlet and the flavor-less state of heavy quark-antiquark bound together by
almost static gluons~\cite{Burnier:2009yu, Dumitru:2009fy, Laine:2006ns}, mainly produce at the very early stages, just after the 
collisions of the ultra-relativistic nuclei and act as an independent degree of freedom. While traversing 
through the medium, they also make transitions to the other quarkonia states with the emission of light hadrons~\cite{Eichten:2007qx}.
Being a bound states of $Q{\bar Q}$ , heavy quarkonia also provide a possibility to explore the important features of quantum
chromodynamics (QCD), the theory of strong interactions, due to the presence of various scales~\cite{Voloshin:2007dx, Patrignani:2012an, Brambilla:2010cs},
in the high temperature.

Quarkonia production has been studied within a number of approaches, namely
the color evaporation model~ \cite{Frawley:2008kk,Berger:2004cc,Amundson:1996qr,Adamczyk:2012ey} 
  which is motivated by the principle of quark-hadron duality, {\it i.e.,}
  it assumes that every produced $c{\bar c}$ evolves into charmonium if it has an
invariant mass less than the threshold for producing a pair of open charm mesons, 
the quarkonia production in color singlet mechanism has been studied in Ref.~\cite{Adamczyk:2012ey, Berger:1980ni, Rapp:2008tf}.
Whereas, the enhancement in the production/suppression of quarkonia through coalescence or the recombination of the quarks/anti-quarks
have been discussed in Ref.~ \cite{Silvestre:2008tw,Scardina:2013nua}.
As the heavy quarks masses, $m_c ~\text{or}~m_b\gg \Lambda_{QCD}$ (QCD scale), the velocity of the bound state of
heavy quarks remain small and hence, the Non-Relativistic QCD (NRQCD) approach~\cite{Bodwin:1994jh,Caswell:1985ui,Pineda:2011dg,Shao:2014yta},
using non-relativistic potential models has also been exploited in the present context.
In this approach, the potential between heavy quarkonia must be approximated by the short distance Coulombic effects
(satisfy asymptotic freedom)  and large distance confinement effects. To that end, the Cornell potential comes as one of
the first possibilities ~\cite{Eichten:1978tg, Eichten:1979ms, Chung:2008sm}, to fulfil these requirements and
describe the interaction between the heavy quark-antiquark pair. 
Recently, the properties of heavy quarkonia have been examined by several authors
~\cite{Luchinsky:2017pby,Fulsom:2017erj,Soto:2017one}. Especially, the production/suppression of quarkonia
has been studied either theoretically or experimentally in Ref.~\cite{Faccioli:2017nfn,
Song:2017phm, TapiaAraya:2017fga, Adare:2011yf, Gale:1999bb, Gale:1998mj, Rosnet:2017iod, Szczurek:2017qul, 
LopezLopez:2017gvp, Hu:2016jcs, Nachtman:2017wmq, Szczurek:2017uvc, Muller:2017csa, Likhoded:2017kfw, Aronson:2017ymv,
Cisek:2017hgs, Thakur:2017kpv},  and the disassociation temperature in Ref.~\cite{Chen:2017jje, Abu-Shady:2017hfo, 
Negash:2017rqt, Braga:2016oem, Ducloue:2016ywt, Vairo:2017hyi, Geiss:1998ma}.

Following our recent work on dissociation of heavy quarkonia, within quasi-particle approach, for the isotropic medium in Ref.~\cite{Agotiya:2016bqr},
the present analysis accommodates the presence of local momentum anisotropy to estimate the dissociation temperature of heavy quarkonia.
While considering the momentum anisotropy, both the oblate and the prolate situation have been taken into account and compared with the isotropic one.
The motivation to incorporate the anisotropy in the study of quarkonia suppression
comes from the fact that, the QGP produced in heavy ion (off-central) collisions does not possess isotropy. Instead, the momentum
anisotropy is present in all the stages of the heavy-ion collisions, and hence, the inclusion of the anisotropy is inevitable.
There are many articles present~\cite{Romatschke:2006bb,Schenke:2006yp, Mauricio:2007vz, Dumitru:2007rp, Baier:2008js, Dumitru:2007hy},
where the impact of the anisotropy in various 
observables of QGP has been investigated. In most of these studies, the ideal Bose/ Fermi distributions~\cite{Carrington:2008sp}, have
been considered in a combination to define the distribution function in isotropic medium. 

Considering the medium as a hot thermal bath instead of non-interacting ideal one, we employed the effective fugacity quasi-particle distribution 
functions to incorporate the hot medium effects, using the effective fugacity quasi-particle model (EQPM)~\cite{chandra_quasi1, chandra_quasi2},
for the isotropic medium. The anisotropy has been introduced at the level of distribution function by stretching and squeezing it in one of the direction,
same as in Ref.~\cite{Romatschke:2003ms,Carrington:2014bla, Jamal:2017dqs, Burnier:2009yu}.
The gluon propagator and in turn, the dielectric permittivity in the presence of anisotropy, in the hot QCD medium has been
obtained using the gluon self-energy. We first calculate the real part and imaginary part of the in-medium Cornell potential,
modified using dielectric permittivity, in the Fourier space. The thermal width and the binding energy of quarkonia 
bound states is then determined by the imaginary and the real part of the modified potential ~\cite{Laine:2006ns, Beraudo:2007ky,
Laine:2007qy,Margotta:2011ta, Strickland:2011aa, Thakur:2013nia}, respectively.
The dissociation temperatures have been calculated by exploiting the criterion ~\cite{Mocsy:2007jz, Burnier07, Escobedo, Laine},
that says, at the dissociation temperature, the thermal width equals twice the (real part of) binding energy.
 To examine the hot QCD medium effects using EQPM ~\cite{chandra_quasi1, chandra_quasi2}, the hot QCD equations of state (EoSs) 
have been updated with the recent lattice~\cite{bazabov2014,fodor2014}, as well as 3-loop HTL perturbative ~\cite{nhaque,
Andersen:2015eoa} calculations. 

The effects of anisotropy will modify the in-medium potential and, in turn, significantly revise the values of dissociation temperature.
In the oblate case, the dissociation temperature has observed to be higher than the isotropic case. While in the prolate case, it is 
observed to be the least among the three cases. 
The tightly bound ground state has higher binding energies and is expected to melt later than the excited state and hence,
they must have a sequential suppression pattern with temperature. The order observed in the present analysis supports the above fact as,
$\Upsilon^{'}$ ($2s$-state of $b{\bar b}$), has been suppressed at smaller temperature than the $\Upsilon$ ($1s$-state of $b{\bar b}$), for all 
 considered EoSs. It has been further seen that the dissociation temperatures using non-ideal EoSs come out smaller as 
compared to the ideal one for each $Q{\bar Q}$-states studied here.

The paper is organized as follows. In section~\ref{HQP}, we shall review heavy-quark potential with its real and imaginary part in the anisotropic medium. 
In this section, we describe the quasi-particle model that has been employed in our analysis along with the discussion of the 
binding energy and melting of heavy quarkonia states. Section~\ref{RD}, refers to the results and discussion part.
In section~\ref{Con}, we shall conclude the present work.

\section{Heavy-quark potential, thermal width and the Quarkonia binding energy in the anisotropic hot QCD medium}
\label{HQP}
The crucial role played by the static heavy-quark potential
to understand the physics behind the quarkonia bound state has been 
studied by the several authors as mentioned earlier. In the present
analysis, we preferred to work with the Cornell potential~\cite{Eichten:1978tg,Eichten:1979ms}, that 
contains the Coulombic as well as the string part given as,
\ba
{\text V(r)} = -\frac{\alpha}{r}+\sigma r,
\label{eq:cor}
\ea
 modifying it in the presence of dissipative medium using the dielectric permittivity, $\epsilon(k)$, in the Fourier space.
Here, $r$ is the effective radius of the corresponding quarkonia state,  $\alpha$ is  the strong coupling constant
and $\sigma$ is the string tension. The modification of the string part along with the Coulombic can 
be exploited due to the fact that, the transition from hadronic phase to the QGP is a crossover ~\cite{Rothkopf:2011db},
and so, the string tension does not vanish abruptly at or near, $T_c$. Let us now briefly discuss the EQPM~\cite{chandra_quasi1, chandra_quasi2},
and then the medium modification of the above potential in the presence of anisotropy will be described. After that, we shall address the binding energy and
thermal width using the derived modified potential.

\subsection{Effective fugacity quasi-particle model(EQPM) and Debye Screening}

EQPM, maps the hot QCD medium effects with the effective equilibrium distribution function, $f_{g,q}(p)$, of quasi-partons
~\cite{chandra_quasi1, chandra_quasi2}, that describes the strong interaction effects in terms of effective fugacities, $z_{g,q}$.
Where the quasi-parton equilibrium distribution functions for gluon and quark/anti-quark, respectively read as,
  \ba
f_{g}(p) = \frac{1}{z_{g}^{-1}e^{\beta E_p} - 1},~~~f_{q/{\bar q}}(p) = \frac{1}{z_{q/{\bar q}}^{-1}~e^{\beta E_p} + 1}.
\label{eq:distr}
  \ea
  
 Using EQPM, the energy dispersion relation modified as,
  \begin{equation*}
 \omega_{g/q,{\bar q}}=E_p+T^2\partial_T \ln(z_{g/q,{\bar q}}),
  \end{equation*}
  
The Debye mass, $m_D$ can be obtained using the distribution functions given in Eq.~\ref{eq:distr} as, 
\ba
  m_D^2&=& -4 \pi \alpha(T) \bigg(2 N_c \int \frac{d^3 p}{(2 \pi)^3} \partial_p f_g ({\bf p})\nn
  &+& 2 N_f  \int \frac{d^3 p}{(2 \pi)^3} \partial_p f_q ({\bf p})\bigg),
  \label{eq:debye}
 \ea
where, $\alpha(T)$ is the running coupling at finite temperature ($T$)~\cite{Laine:2005ai}. $N_c$ and $N_f$ are the color degrees of freedom and 
the number of flavor, respectively. Applying quasi-parton equilibrium distribution function from Eq.~\ref{eq:distr} in Eq.~\ref{eq:debye}, we have, 
 \ba
 {m_D^2}^{(EoS(i))}&=&4 \pi \alpha(T) T^2  \bigg( \frac{2 N_c}{\pi^2} PolyLog[2,z_g^i]\nn
 &-&\frac{2 N_f}{\pi^2} PolyLog[2,-z_q^i]\bigg). 
 \ea
Where the index-$i$, denotes the different EoSs, incorporating the QCD interactions modeled from improved perturbative 3-loop HTL QCD computations
by N. Haque {\it et, al.}~\cite{nhaque, Andersen:2015eoa} and recent $(2+1)$-flavor lattice QCD simulations~\cite{bazabov2014,fodor2014}.

In view of the fact that weak perturbative(resummed) computations on the Equation of State (EoS) in hot QCD show nice convergence properties
and agree well with the lattice QCD results. The strong interaction effects encoded in Lattice EoS (LEoS) could be applied to effective gluonic
and quark/anti-quark degrees of freedom and utilize to develop effective transport theory in those regions where weak perturbative results make 
sense, and transport theory could lead to reliable outcomes. The above work is done more in the above-mentioned spirit. In other words, with
EQPM for LEoS, we can not go much closer to $T_c$, the analysis is reliable beyond $T_c$  ($T \gtrsim T_c$). Both EQPM and effective 
(linearized) transport theory methods will not work at very close to $T_c$. However, in the above-mentioned temperature, these methods could be
used to take care of the interaction in an effective way.  Working at the temperature, $T = 3~T_c$, we have studied the interaction effects
which are important and were not considered in work done earlier in this field.

In the  limit, $z_{g,q}\to 1$ the $m_D$, reduces to the leading order (LO) or ideal case given as,
 \ba
 {m_D^2}^{(LO)}= 4\pi\alpha(T) \ T^2 (\frac{N_c}{3}+\frac{N_f}{6}). 
 \ea
Let us now discuss the modification of the potential, considering the presence of anisotropy in the hot QCD medium.
 
 \subsection{Medium modified heavy-quark potential in the presence of anisotropy.}

Here, the anisotropy is introduced due to the fact that, in the off-central relativistic heavy-ion collisions
the spatial anisotropy generates at the very primary stages. As the system evolves with time
different pressure gradients produce in different directions which maps the spatial anisotropy to the momentum anisotropy.
The anisotropy in the present formalism has been introduced at the particle phase space distribution level. Employing the method used 
in Ref.~\cite{Romatschke:2003ms, Carrington:2014bla, Jamal:2017dqs}, the anisotropic distribution functions
has been obtained from isotropic one by rescaling (stretching and squeezing)
it in one of the direction in the momentum space as,
\ba
f({\mathbf{p}})\rightarrow f_{\xi}({\mathbf{p}}) = C_{\xi}~f(\sqrt{{\bf p}^{2} + \xi({\bf p}\cdot{\bf \hat{n}})^{2}}),
\label{aniso_distr}
\ea

where $f({\mathbf{p}})$, is effective fugacity quasi-particle distribution function for the isotropic 
medium~\cite{chandra_quasi1, chandra_quasi2}.The ${\mathbf{\hat{n}}}$, is a unit 
vector (${\bf \hat{n}}^{2} = 1$), showing the direction of momentum anisotropy.  The parameter $\xi$, gives the anisotropic strength in the medium,
and describes the amount of squeezing ($\xi > 0$, or oblate form) and stretching ($-1<\xi<0$, or prolate form) in the ${\bf \hat{n}}$, direction. 
Since the EoSs effects enter through the Debye mass ($m_D$), we want to make it intact from the effects of anisotropy present
in the medium so that it remains same in both the mediums (isotropic and anisotropic), as done in Ref.~\cite{Carrington:2014bla}.
Doing so, only the effects of different EoSs will be carried in the $m_D$, and hence, the normalization constant, $C_{\xi}$ comes out as,

\ba
  C_{\xi}= 
\begin{cases}
   \frac{\sqrt{|\xi|}}{\tanh^{-1}\sqrt{|\xi|}}& \text{if }~~ -1\leq\xi<0\\
    \frac{\sqrt{\xi}}{\tan^{-1}\sqrt{\xi}}    & \text{if }~~ \xi\geq0.
\end{cases}
\label{aniso_const}
\ea
In the small $\xi$ limit, we have

\ba
  C_{\xi}= 
\begin{cases}
   1-\frac{\xi}{3} +O\left(\xi ^{\frac{3}{2}}\right)& \text{if }~~ -1\leq\xi<0\\
    1+\frac{\xi}{3} +O\left(\xi ^{\frac{3}{2}}\right)   & \text{if }~~ \xi\geq0.
\end{cases}
\label{aniso_const}
\ea

 To modify the potential due to the presence of dissipative anisotropic hot QCD medium, the assumption given
 in Ref.~\cite{Agotiya:2008ie} has been followed which says that, the in-medium modification can be obtained
 in the Fourier space by dividing the heavy-quark potential from the medium dielectric permittivity, $\epsilon({\bf k})$ as,
 \ba
 \grave{V}(k)=\frac{{\bar{\text V}}(k)}{\epsilon(k)}.
 \ea
 By making the inverse Fourier transform, we can obtain the modified (or in-medium corrected) potential as,
 \ba 
V(r)= \int \frac{d^3\mathbf{k}}{(2\pi)^{3/2}}(e^{i\mathbf{k} \cdot \mathbf{r}}-1)\grave{V}(k),
 \label{eq:V}
 \ea
 where ${\bar{\text V}}(k)$, is the Fourier transform of ${\text V(r)}$, shown in Eq.~\ref{eq:cor}, given as,
\ba
{\bar{\text V}}(k)= -\sqrt\frac{2}{\pi}\bigg(\frac{\alpha}{k^2}+2\frac{ \sigma}{k^4}\bigg).
\ea

Now, to modify the potential, we first need to calculate the dielectric permittivity which obtains from the self-energy using finite temperature QCD.
It is important to note that the perturbative theory at $T>0$ suffers from the infrared singularities and gauge dependent
results because the perturbative expansion is incomplete at $T>0$. There are infinitely many higher order diagrams with
more and more loops that can contribute to lower order in the coupling constant ~\cite{Thoma:2000dc}. This problem can
be partly avoided by using the HTL resummation technique ~\cite{Braaten:1989mz} and one can obtain the consistent results
up to the leading order.
Another equivalent approaches to obtain $\epsilon({\bf k})$, is the many-particle kinetic theory (or the semi-classical transport theory)
which provides the same results up to one-loop order (or in the Abelian limit)~\cite{Blaizot:1993be, Kelly:1994dh, Jamal:2017dqs}.
  Exploiting any of these two methods, one finds the gluon self-energy, $\Pi^{\mu\nu}$, and then the
   static gluon propagator that represents the inelastic scattering of an off-shell gluon to a thermal gluon as,
    
 \ba
 \Delta^{\mu\nu}(\omega,{\bf k}) = k^{2}g^{\mu\nu} - k^{\mu} k^{\nu} + \Pi^{\mu\nu}(\omega,{\bf k}).
 \ea
 Next, the dielectric tensor can be obtained in the static limit, in the Fourier space, from the temporal component 
 of the propagator as,
 \ba
\epsilon^{-1}({\bf k}) = -\lim_{\omega \to 0}k^2\Delta^{00}(\omega,{\bf k}).
\label{eq:eps}
\ea
 
 Now, to obtain the real part of the inter-quark potential in the static limit, the temporal component 
 of  real part of the retarded (or advanced) propagator in the Fourier space is demanded, which is given as
  \ba
  Re[\Delta^{00}_{R(A)}]({\omega = 0,\bf k}) &=&\frac{-1}{k^2+m_{D}^2}-\xi\Big(\frac{1}{3(k^2+m_{D}^2)}\nn
  &-&  \frac{m_{D}^2(3\cos{2\theta_n}-1)}{6(k^2+m_{D}^2)^2}\Big).
\label{eq:Re_delta}
\ea

The imaginary part of the same can be derived from the imaginary part of the temporal component of symmetric propagator
in the static limit. That can be seen as,
\ba
Im[\Delta^{00}_{S}]({\omega = 0,\bf k})& =& \pi~ T~ m_{D}^2\bigg(\frac{-1}{k(k^2+m_{D}^2)^2}\nn 
&+&\xi\Big(\frac{-1}{3k(k^2+m_{D}^2)^2}+\frac{3\sin^{2}{\theta_n}}{4k(k^2+m_{D}^2)^2}\nn
&-&\frac{2m_{D}^2\big(3\sin^{2}({\theta_n})-1\big)}{3k(k^2+m_{D}^2)^3}\Big)\bigg),
\label{eq:Im_delta}
\ea
where
\ba
\cos(\theta_n)=\cos(\theta_r)\cos(\theta_{pr})+ \sin(\theta_r)\sin(\theta_{pr})\cos (\phi_{pr}).\nn
\ea

In the above expression the angle $\theta_n$, is in between the particle momentum ${\bf p}$, and the direction
of anisotropy, ${\bf {\hat n}}$. The angle between ${\bf r}$, and ${\bf n}$, is $\theta_r$. $\phi_{pr}$, and
$\theta_{pr}$, are respectively, the azimuthal and the polar angle between ${\bf p}$ and ${\bf r}$.
Next, to modify the real part of the potential, $\epsilon({\bf k})$ can be obtained using Eq.\ref{eq:Re_delta} in Eq.\ref{eq:eps} as,
\ba
\epsilon^{-1}({\bf k})&=&\frac{k^2}{k^2+m_{D}^2}+k^2\xi\Big(\frac{1}{3(k^2+m_{D}^2)}\nn &-&  \frac{m_{D}^2(3\cos{2\theta_n}-1)}{6(k^2+m_{D}^2)^2}\Big).
\label{eq:Re_eps}
\ea

Similarly, the imaginary part of the potential can be modified by using,  $\epsilon({\bf k})$ which can be obtained by employing Eq.\ref{eq:Im_delta} in Eq.\ref{eq:eps} as,
\ba
\epsilon^{-1}({\bf k})& =& \pi~ T~ m_{D}^2\bigg(\frac{k^2}{k(k^2+m_{D}^2)^2}-\xi k^2\Big(\frac{-1}{3k(k^2+m_{D}^2)^2}\nn
&+&\frac{3\sin^{2}{\theta_n}}{4k(k^2+m_{D}^2)^2}-\frac{2m_{D}^2\big(3\sin^{2}({\theta_n})-1\big)}{3k(k^2+m_{D}^2)^3}\Big)\bigg).\nn
\label{eq:Im_eps}
\ea

In the limit, $T\rightarrow0$, and in the absence of anisotropy, the real part of the $\epsilon^{-1}(k)$ goes to unity  while the imaginary part vanishes
and thus, the modified potential simply reduces to the Cornell form.
Let us now discuss, the real and the imaginary potential, modified using the above define $\epsilon^{-1}({\bf k})$, separately in the next two sub-sections.

\subsubsection{Real part of the potential in the anisotropic medium}
Using Eq.\ref{eq:Re_eps} in Eq.\ref{eq:V}, we can write the real part of the potential as,
 \ba
Re[V({\bf r},\xi,T)]&=&\int \frac{d^3\mathbf{k}}{(2\pi)^{3/2}}(e^{i\mathbf{k} \cdot \mathbf{r}}-1)\bigg(-\sqrt{\frac{2}{\pi}}\frac{\alpha}{k^2}\nn
&-&\frac{4\sigma}{\sqrt{2\pi}k^4}\bigg)\bigg(\frac{k^2}{k^2+m_{D}^2}+k^2\xi\Big(\frac{1}{3(k^2+m_{D}^2)}\nn
&-& \frac{m_{D}^2(3\cos{2\theta_n}-1)}{6(k^2+m_{D}^2)^2}\Big)\bigg).\nn
\ea

Solving the above integral, we find 
 \ba
  Re[V({\bf r},\xi,T)] &=&\alpha ~ m_D \left(-\frac{e^{-s}}{s}-1\right)+\frac{\sigma}{ m_D}  \Big(\frac{2 e^{-s}}{s}\nn
  &-&\frac{2}{s}+2\Big)+\alpha ~ \xi ~ m_D \bigg[-\frac{3 \cos (2 \theta_r )}{2 s^3}\nn
  &-&\frac{1}{2 s^3}+\frac{1}{6}+e^{-s} \Big\{\frac{1}{2 s^3}+\frac{1}{2 s^2}+\Big(\frac{3}{2 s^3}\nn
  &+&\frac{3}{2 s^2}+\frac{3}{4 s}+\frac{1}{4}\Big) \cos (2 \theta_r )+\frac{1}{4 s}-\frac{1}{12}\Big\}\bigg]\nn
  &+&\frac{\xi ~ \sigma}{m_D}  \bigg[\left(\frac{6}{s^3}-\frac{1}{2 s}\right) \cos (2 \theta_r )+\frac{2}{s^3}-\frac{5}{6 s}\nn
  &+&\frac{1}{3}+e^{-s} \Big\{-\frac{2}{s^3}-\frac{2}{s^2}+\Big(-\frac{6}{s^3}-\frac{6}{s^2}\nn
  &-&\frac{5}{2 s}-\frac{1}{2}\Big) \cos (2 \theta_r )-\frac{1}{6 s}+\frac{1}{6}\Big\}\bigg]
  \label{eq:RV}
\ea
where $s = r m_D$. Considering the limit, $s\ll1$ in Eq.\ref{eq:RV} we have,
\ba
Re[V({\bf r},\xi,T)] &=&\frac{ s~ \sigma }{m_D}\left(1+\frac{\xi }{3}\right)-\frac{\alpha~  m_D}{s} \bigg(1+\frac{s^2}{2}\nn 
&+&\xi  \left(\frac{1}{3}+\frac{s^2}{16}\left(\frac{1}{3}+ \cos \left(2 \theta _r\right)\right)\right)\bigg).
\label{eq:realv}
\ea

Here, in the isotropic limit, one can observe that there is an additional term in $s$ with $\alpha$ in Eq.\ref{eq:realv}.
This term vanishes in the limit, $T\rightarrow 0$ and we end up with the vacuum potential while it contributes as a thermal correction to the real part of the medium modifed potential at $T\neq0$. 

\begin{figure*}   
    \includegraphics[height=7cm,width=8.6cm]{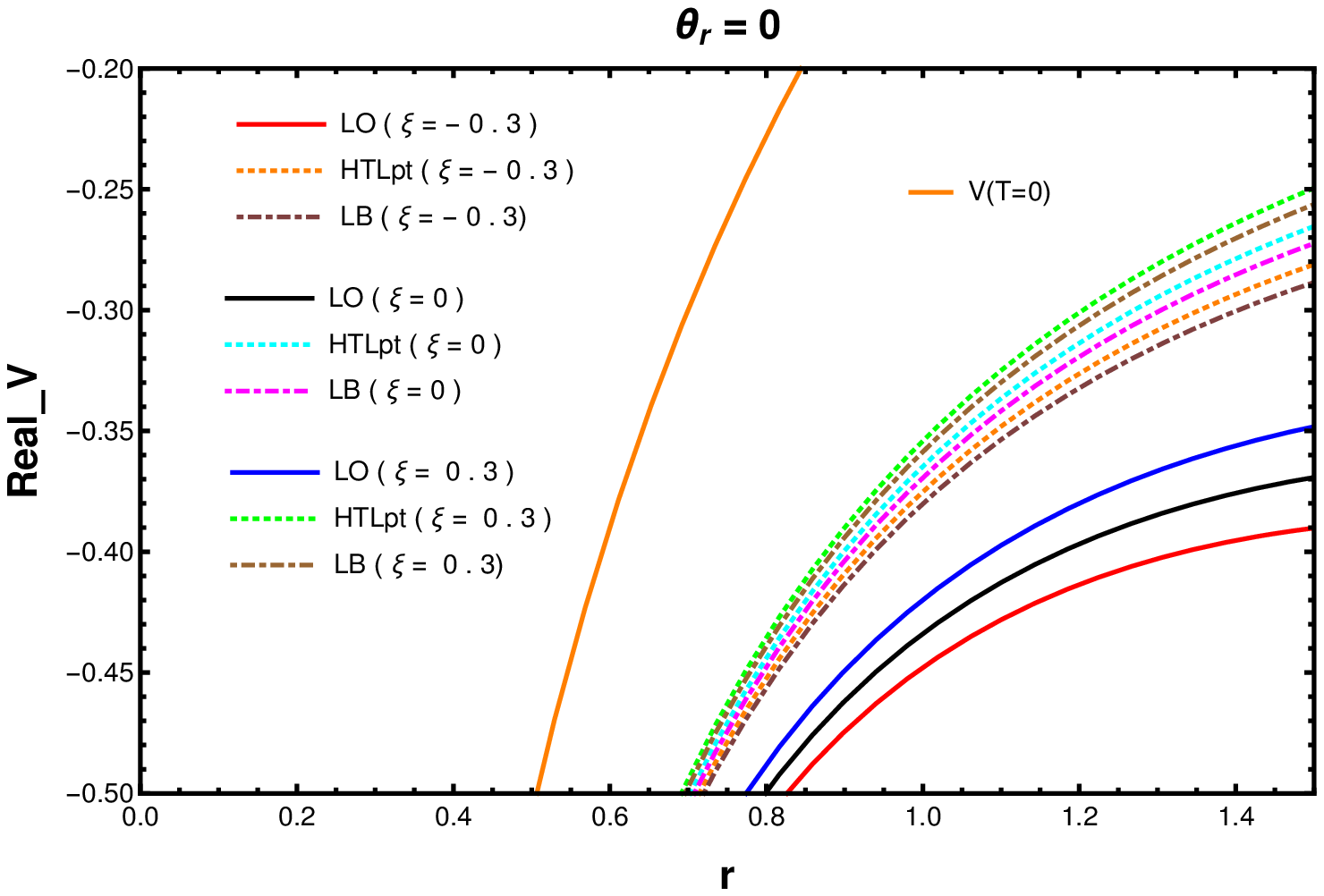}
    \hspace{1mm}
    \includegraphics[height=7cm,width=8.6cm]{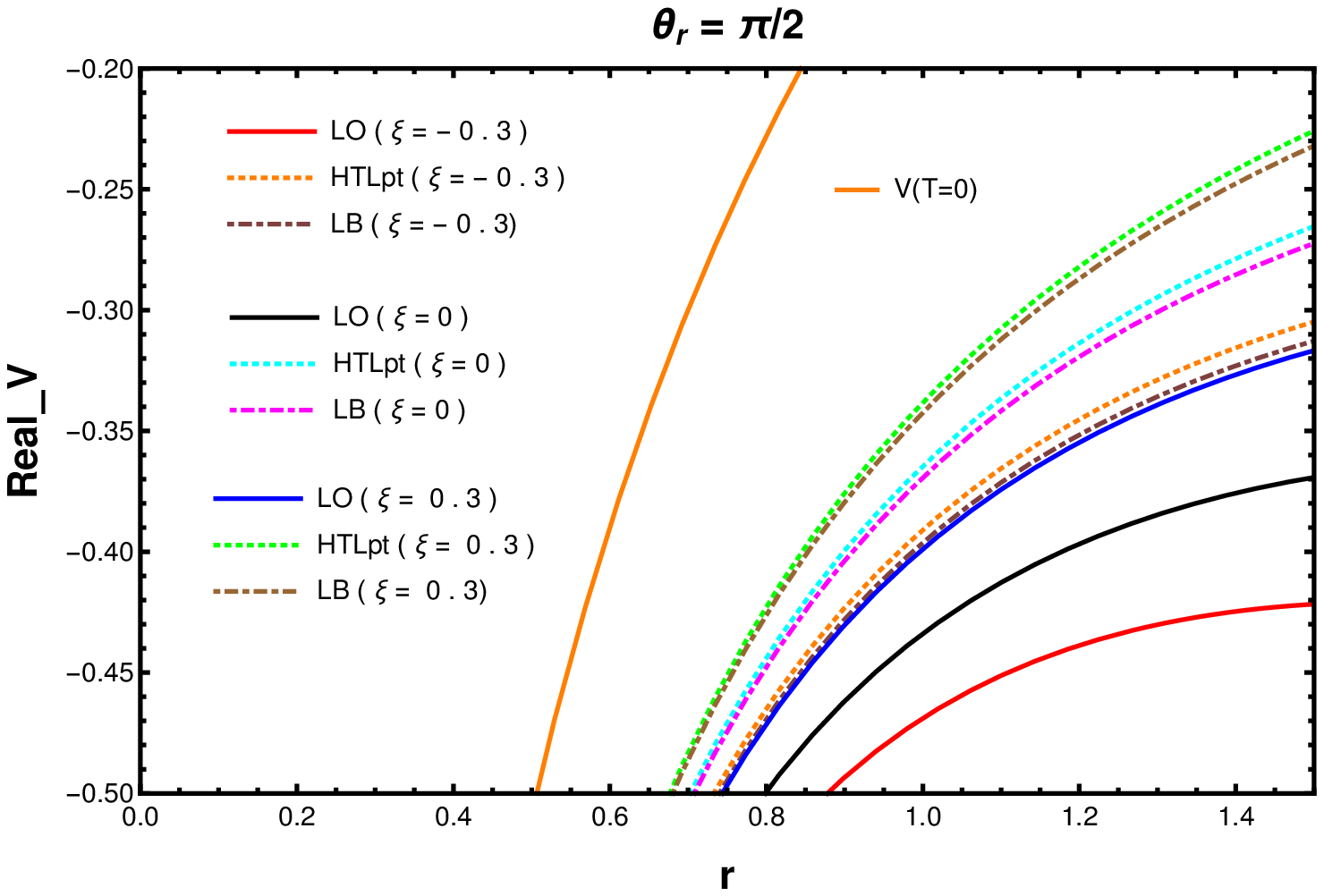}
    \caption{Real part of the medium modified potential for $\theta_r = 0$ (left panel) and $\theta_r = \pi/2$ (right panel) 
    with various EOSs and different, $\xi$ at fixed $T_{c} = 0.17~$GeV and $T = 3~T_c ~$GeV along with the potential at $T=0$.}
    \label{fig:realV}
\end{figure*}

\begin{figure*}   
    \includegraphics[height=7cm,width=8.6cm]{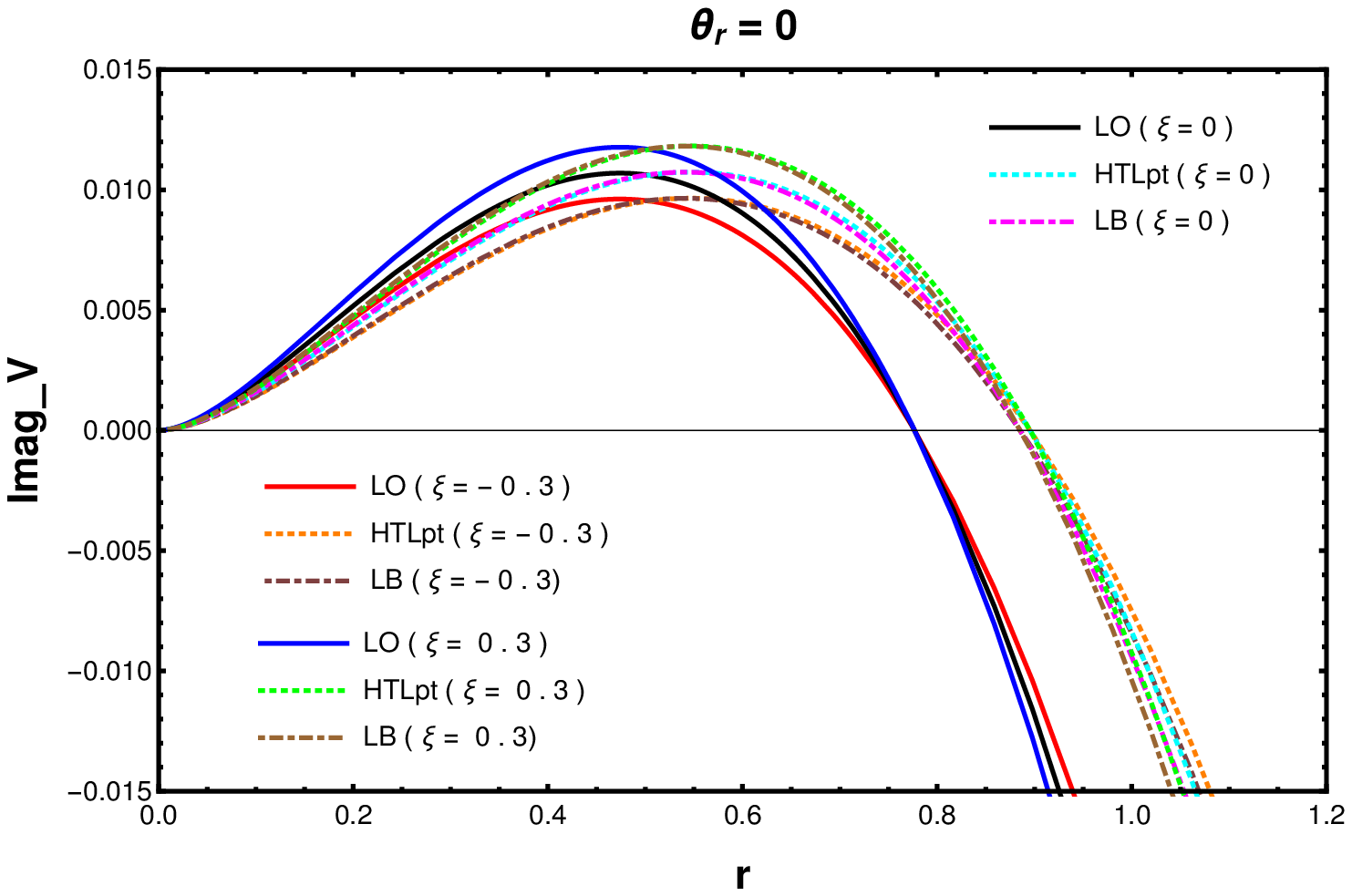}
    \hspace{1mm}
    \includegraphics[height=7cm,width=8.6cm]{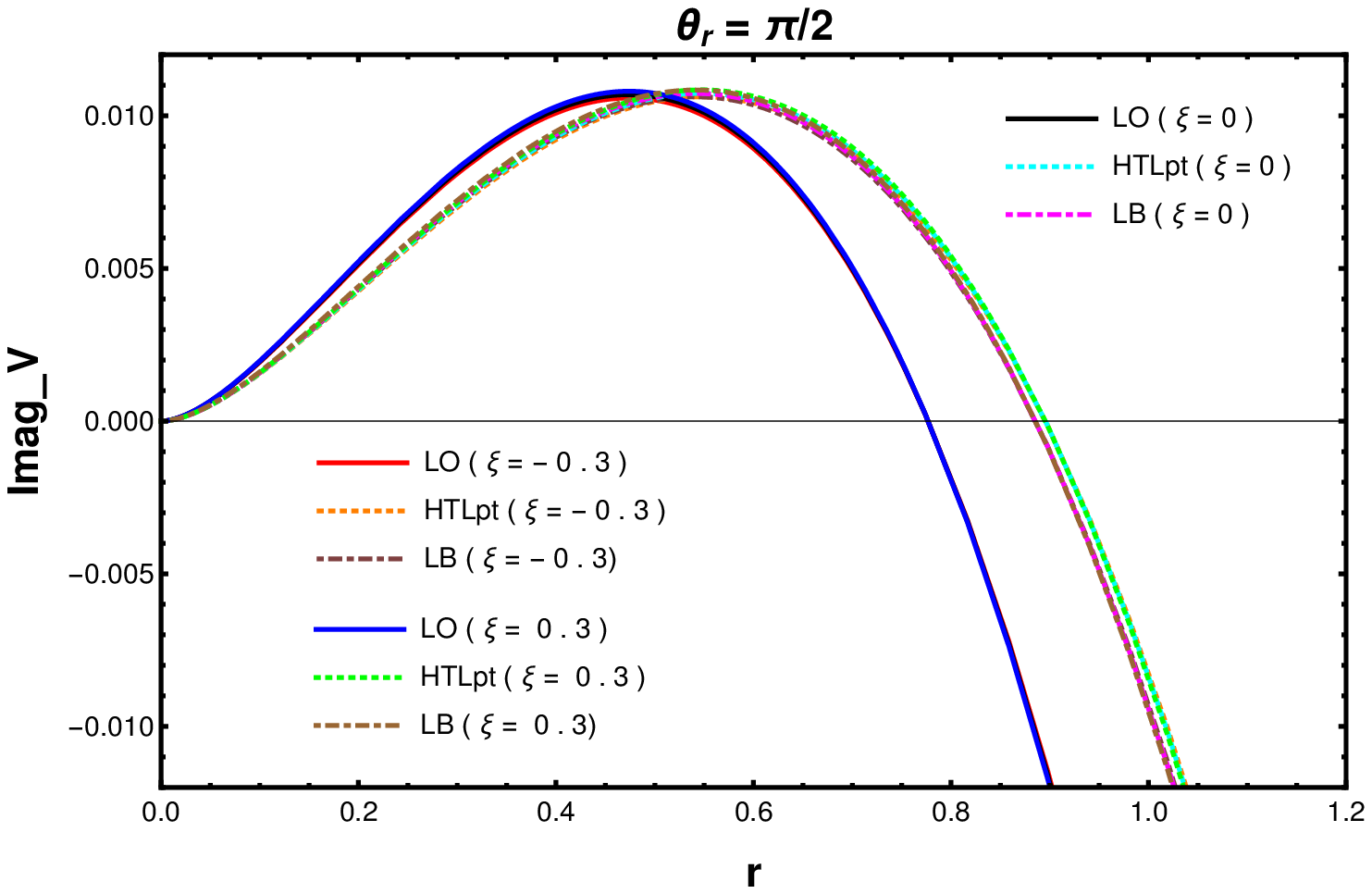}
    \caption{Imaginary part of the medium modified potential for $\theta_r = 0$ (left panel) and $\theta_r = \pi/2$ (right panel) 
    with various EOSs and different, $\xi$ at fixed $T_{c} = 0.17~$GeV and $T = 3~T_c ~$GeV.}
    \label{fig:imagV}
\end{figure*}

\subsubsection{Imaginary part of the potential in the anisotropic medium}
The imaginary potential, using Eq.\ref{eq:Im_eps} in Eq.\ref{eq:V}, can be written as,
\ba
\text{Im}[V({\bf r},\xi,T)]&=&\pi~ T~ m_{D}^2\int \frac{d^3\mathbf{k}}{(2\pi)^{3/2}}(e^{i\mathbf{k} \cdot \mathbf{r}}-1)\bigg(-\sqrt{\frac{2}{\pi}}\frac{\alpha}{k^2}\nn
&-&\frac{4\sigma}{\sqrt{2\pi}k^4}\bigg)\bigg(\frac{k}{(k^2+m_{D}^2)^2}-\xi\Big(\frac{-k}{3(k^2+m_{D}^2)^2}\nn
&+&\frac{3k\sin^{2}{\theta_n}}{4(k^2+m_{D}^2)^2}-\frac{2m_{D}^2 k\big(3\sin^{2}({\theta_n})-1\big)}{3(k^2+m_{D}^2)^3}\Big)\bigg).\nn
\ea

To solve the above equation, we separate Coulombic term (containing, $\alpha$) and the string term (having, $\sigma$)  as,
\ba
\text{Im}[V({\bf r},\xi,T)]= \text{Im} V_{1} ({\bf r},\xi,T)+ \text{Im} V_{2} ({\bf r},\xi,T).
\ea

 \ba
 \text{Im} V_{1} ({\bf r},\xi,T)&=&\frac{\alpha ~T ~m^2_D}{2~\pi}\int d^3{\bf k}(e^{i\mathbf{k} \cdot \mathbf{r}}-1)
 \frac{1}{k}\bigg[\frac{-1}{(k^2+m_D^2)^2}\nn
 &+&\xi[\frac{-1}{3(k^2+m_D^2)^2}+\frac{3\sin^2\theta_n}{2(k^2+m_D^2)^2}\nn
 &-&\frac{4 m_D^2(\sin^2\theta_n-\frac{1}{3})}{(k^2+m_D^2)^3}]\bigg]
 \label{eq:im_v1}
\ea
and
  \ba
\text{Im} V_{2} ({\bf r},\xi,T)&=&\frac{\sigma~ T~ m^2_D}{\pi}\int d^3\mathbf{k}(e^{i\mathbf{k} \cdot \mathbf{r}}-1)
\frac{1}{k^3}\bigg[\frac{-1}{(k^2+m_D^2)^2}\nn
&+&\xi[\frac{-1}{3(k^2+m_D^2)^2}+\frac{3\sin^2\theta_n}{2(k^2+m_D^2)^2}\nn
&-&\frac{4 m_D^2(\sin^2\theta_n-\frac{1}{3})}{(k^2+m_D^2)^3}]\bigg].
 \label{eq:im_v2}
 \ea
 
The contribution due to the Coulombic part in the imaginary potential considering the limit, $r m_D\equiv s\ll 1$ is found to be,

 \ba
\text{Im} V_{1} (r,\theta_r,T)&=&-\frac{ \alpha~ s^2~ T}{180}\Big\{\xi  \big(9 \cos2 \theta_r-7\big)\nn&+&60\Big\}\log \left(\frac{1}{s}\right)
\ea 
and from the string part we obtained,
\ba
\text{Im} V_{2} (r,\theta_r,T)&=&-\frac{s^4 ~\sigma ~ T }{1260 ~m_D^2}\Big\{\xi  \big(9 \cos2 \theta_r-4\big)\nn
&+&42\Big\}\log \left(\frac{1}{s}\right),
\ea 

Hence, the imaginary part of the modified potential in the anisotropic medium is given as,

\ba
\text{Im}[V (r,\theta_r,T)]&=&\frac{\alpha ~ s^2~ T}{3} \Big\{\frac{ \xi }{60} (7-9 \cos 2 \theta_r)-1\Big\}\log \left(\frac{1}{s}\right)\nn
&+&\frac{s^4 ~\sigma ~ T}{m_D^2}\Big\{\frac{\xi}{35}  \left(\frac{1}{9}-\frac{1}{4} \cos 2 \theta_r \right)\nn
&-&\frac{1}{30}\Big\}\log \left(\frac{1}{s}\right).
 \label{eq:imagv}
 \ea

\begin{figure*}   
    \includegraphics[height=5cm,width=5.8cm]{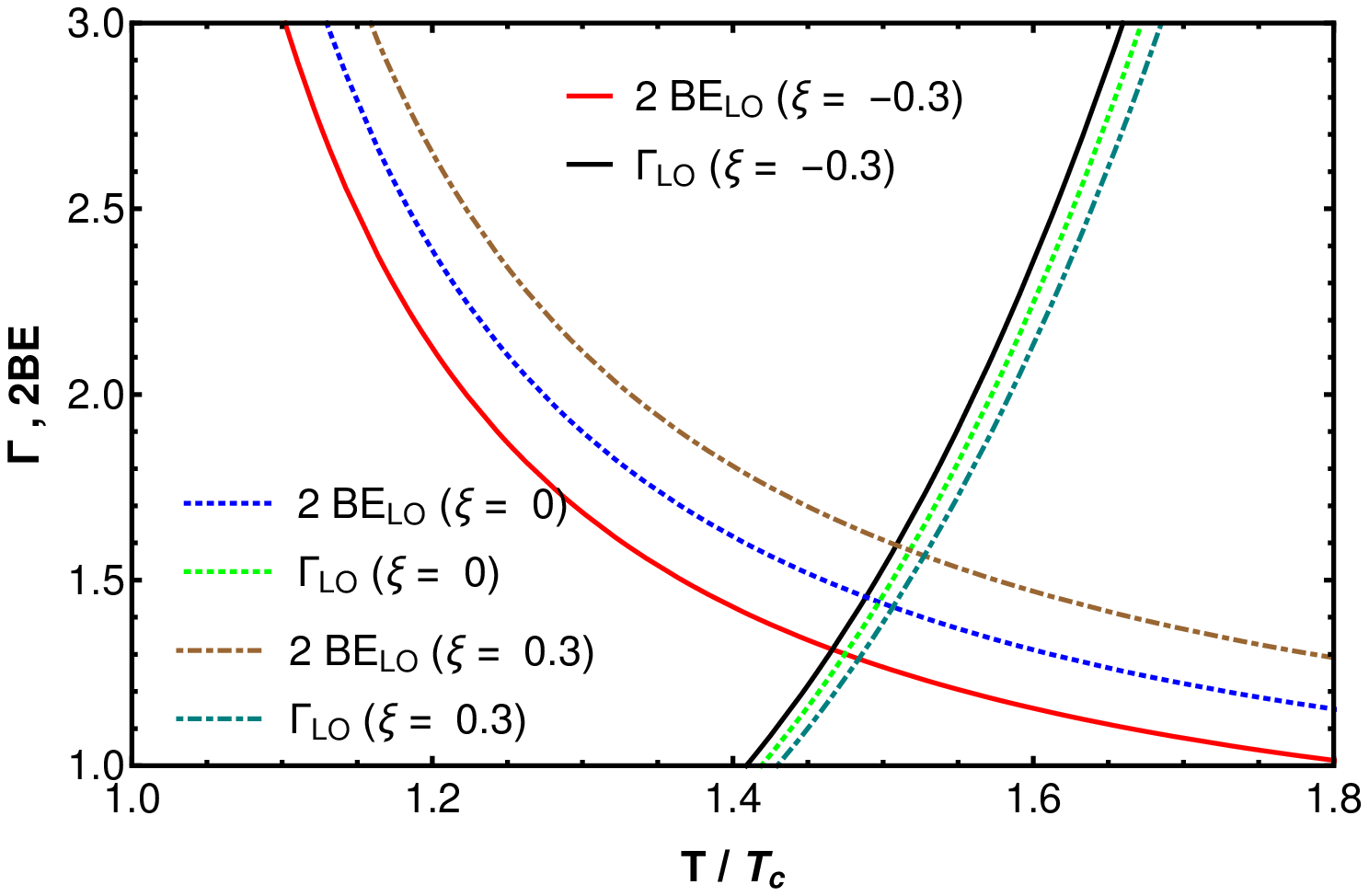}
    \hspace{-1mm}
    \includegraphics[height=5cm,width=5.8cm]{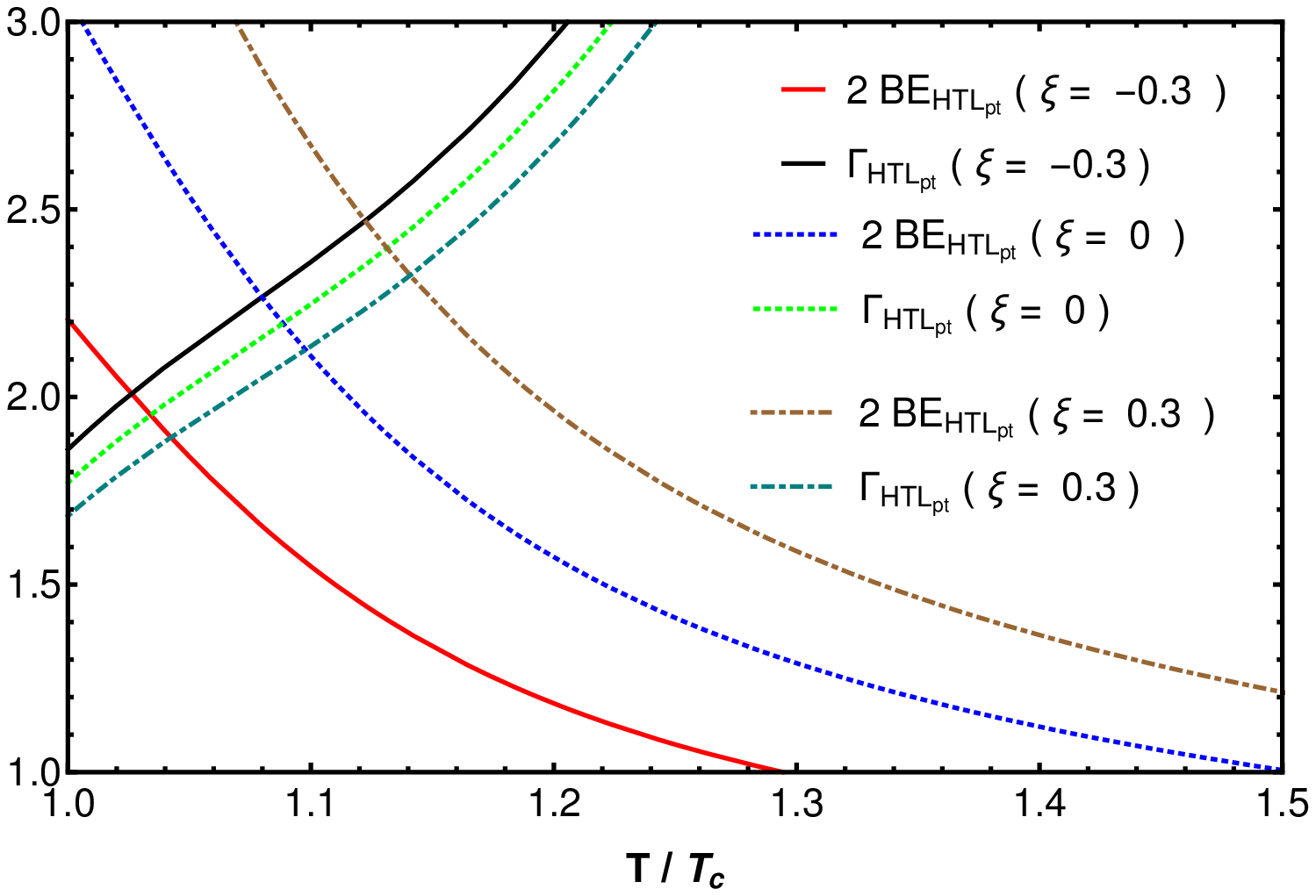}
    \hspace{-1mm}
    \includegraphics[height=5cm,width=5.8cm]{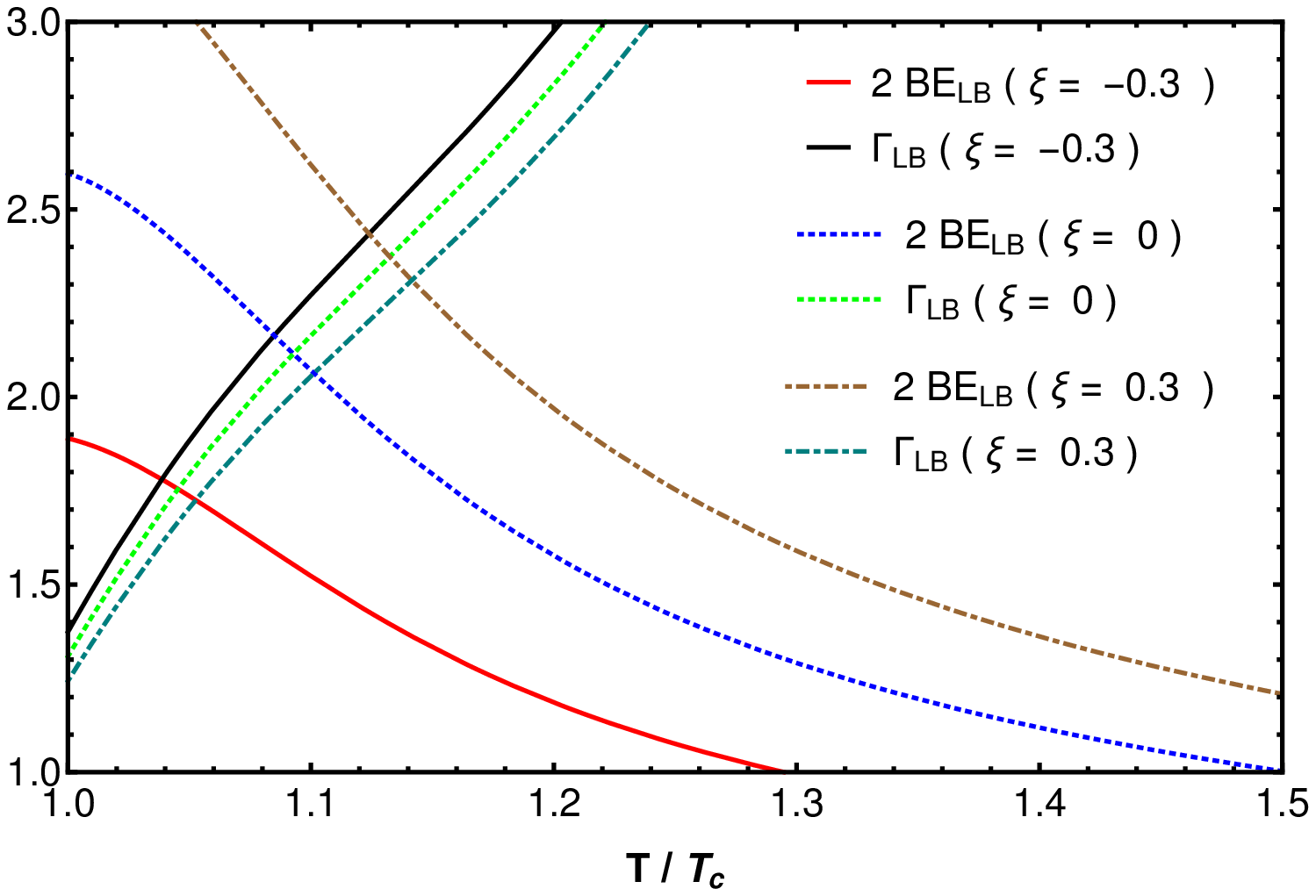}
    \caption{$\Gamma$, 2BE($E_b$) vs $T/T_c$ for $J/\psi$ at $T_{c} = 0.17GeV$ with different $\xi$. We have plotted 
    the leading order (non-interacting) results (left panel) along with the 3-loop HTLpt (middle panel) and Lattice (right panel).}
    \label{fig:jpsi}
\end{figure*}
\begin{figure*}   
    \includegraphics[height=5cm,width=5.8cm]{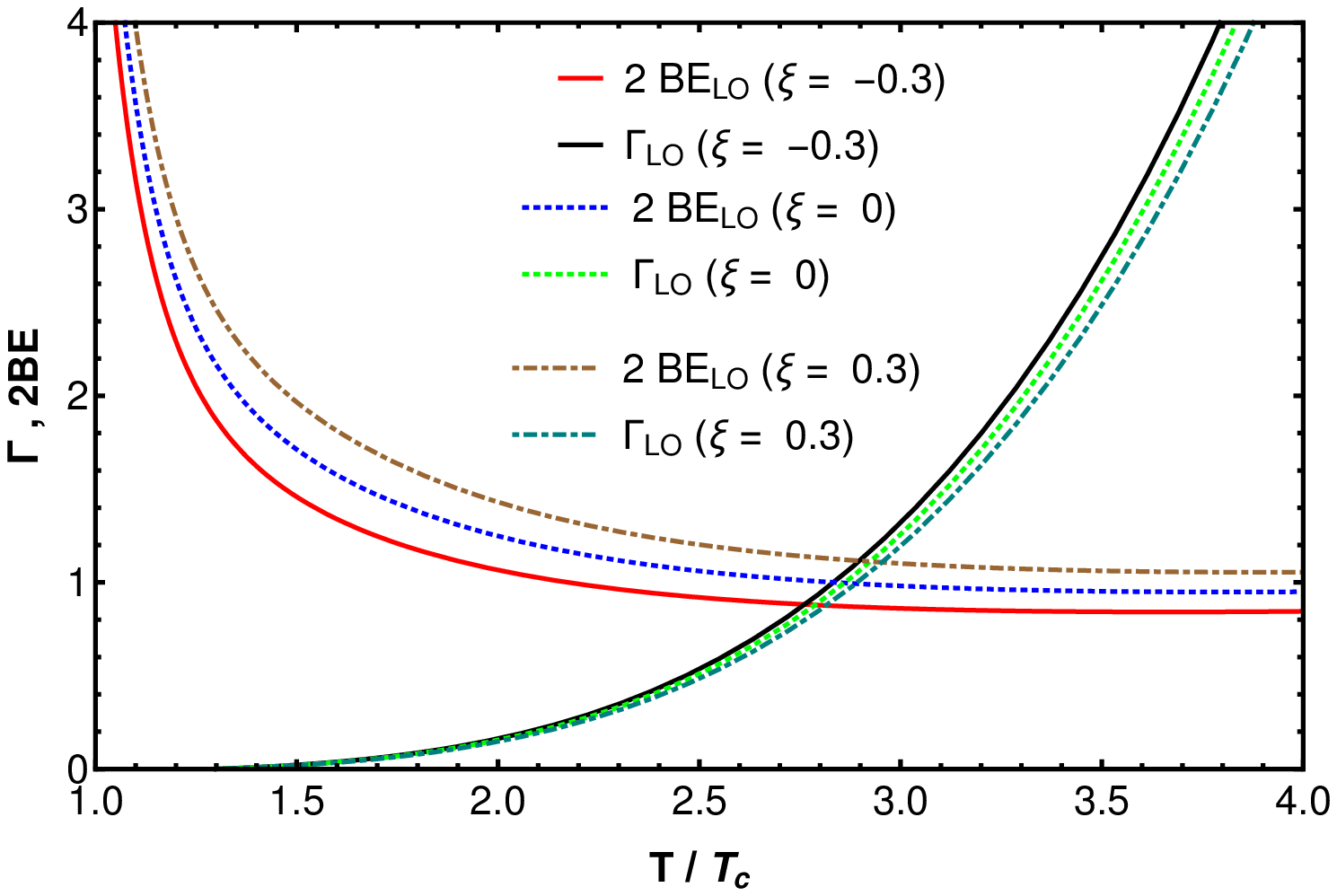}
    \hspace{-1mm}
    \includegraphics[height=5cm,width=5.8cm]{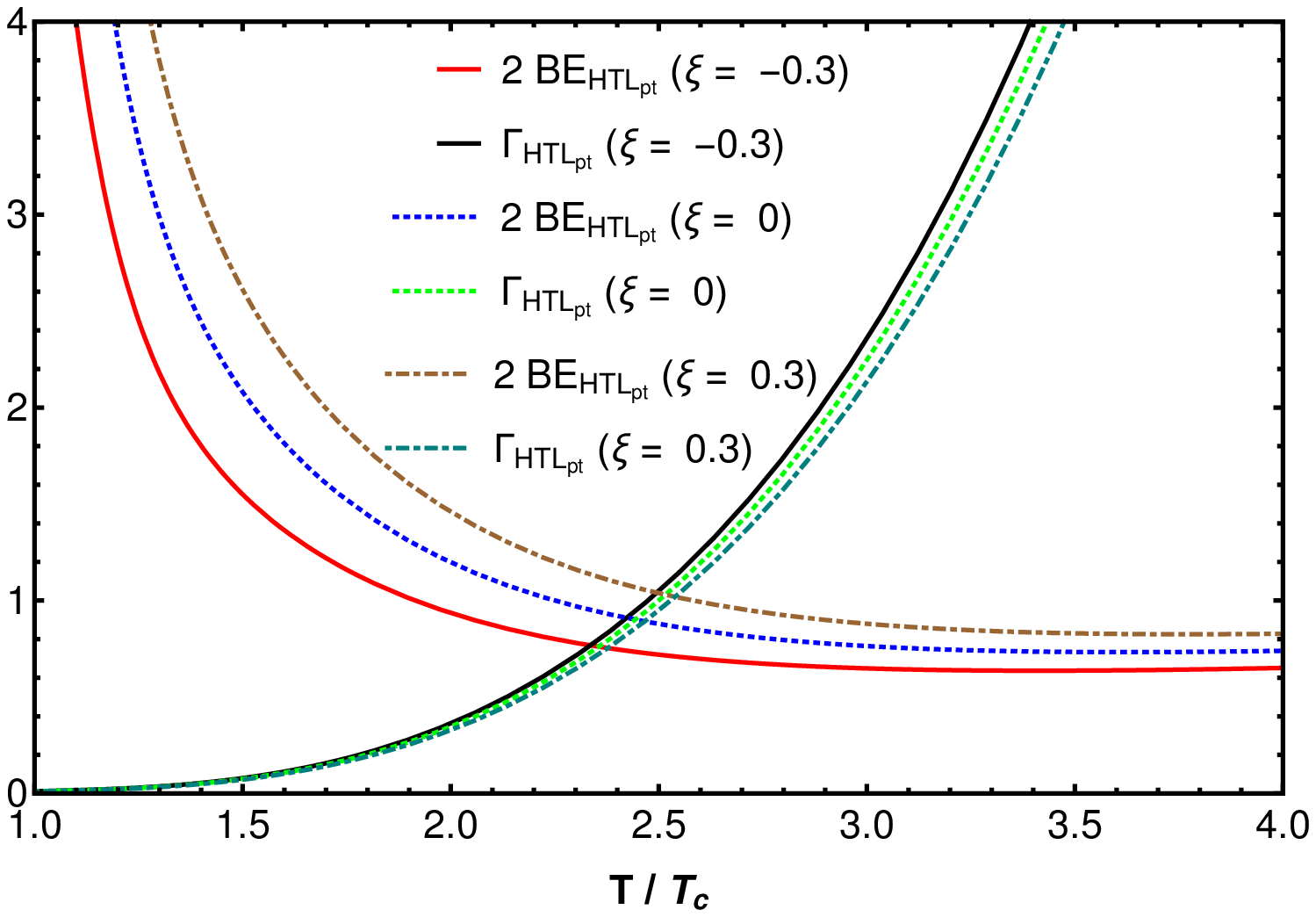}
    \hspace{-1mm}
    \includegraphics[height=5cm,width=5.8cm]{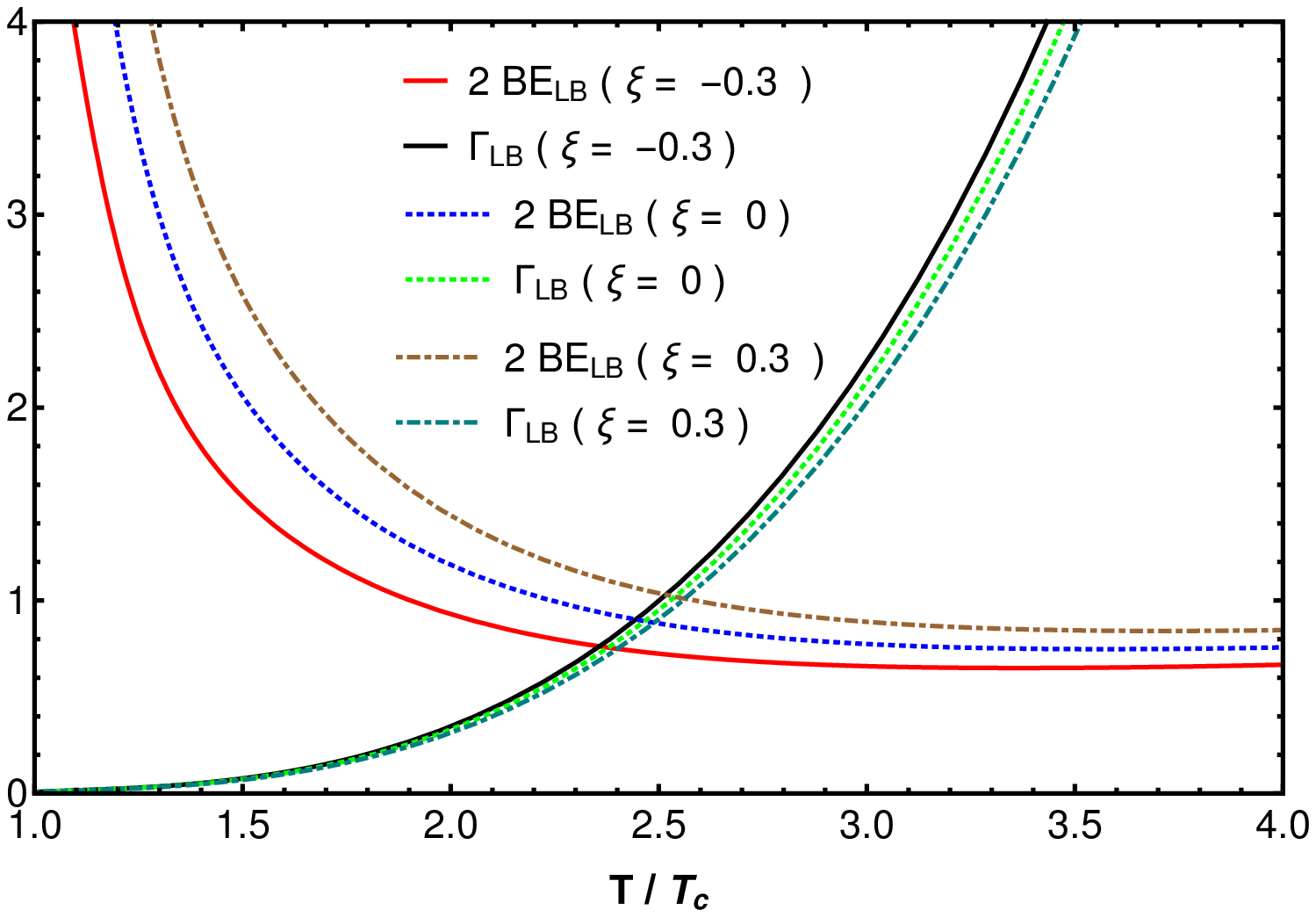}
   \caption{$\Gamma$, 2BE($E_b$) vs $T/T_c$ for $\Upsilon$ at $T_{c} = 0.17GeV$ with different $\xi$. We have plotted 
    the leading order (non-interacting) results (left panel) along with the 3-loop HTLpt (middle panel) and Lattice (right panel).}
    \label{fig:upsilon}
\end{figure*}
\begin{figure*}   
    \includegraphics[height=5cm,width=5.8cm]{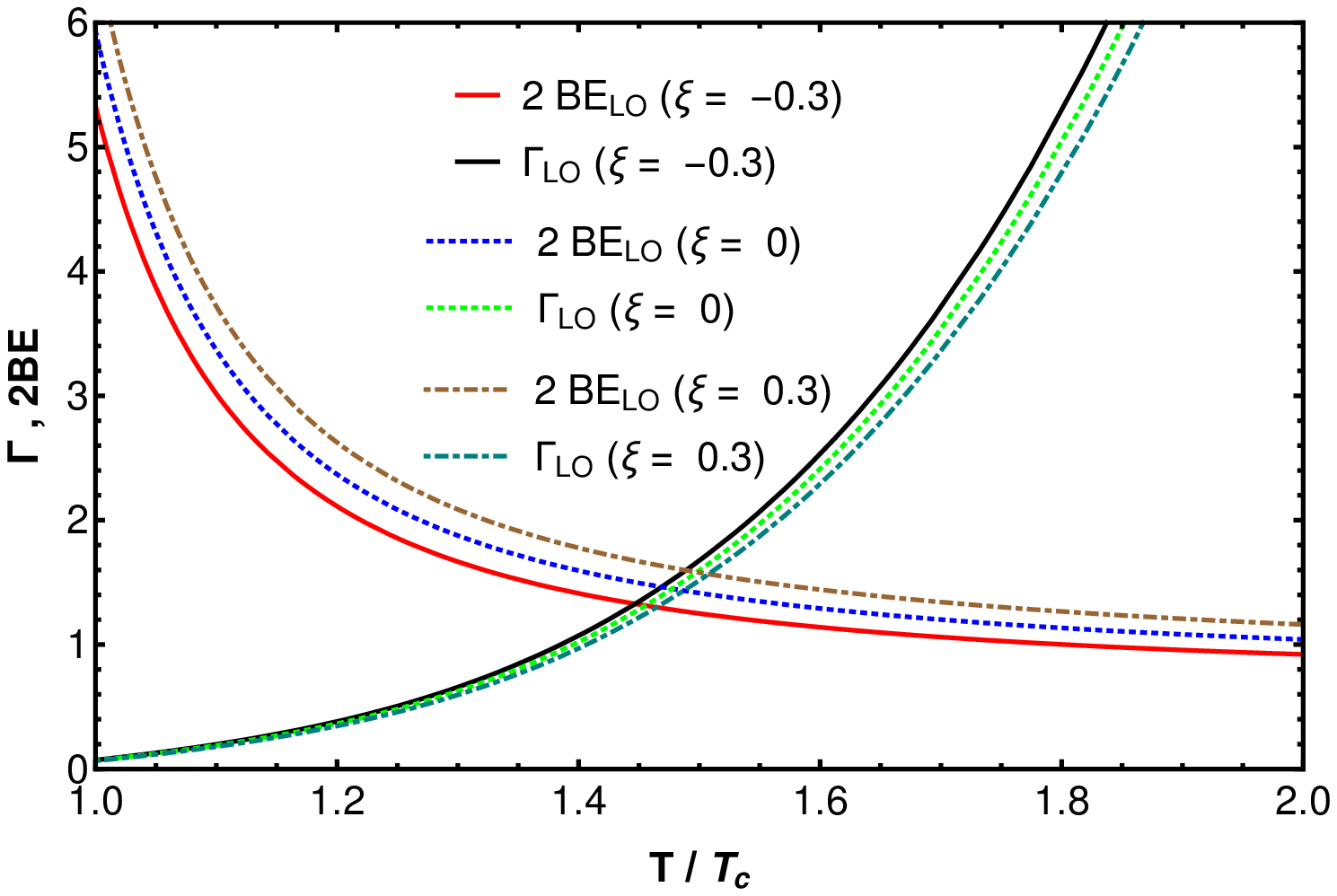}
    \hspace{-1mm}
    \includegraphics[height=5cm,width=5.8cm]{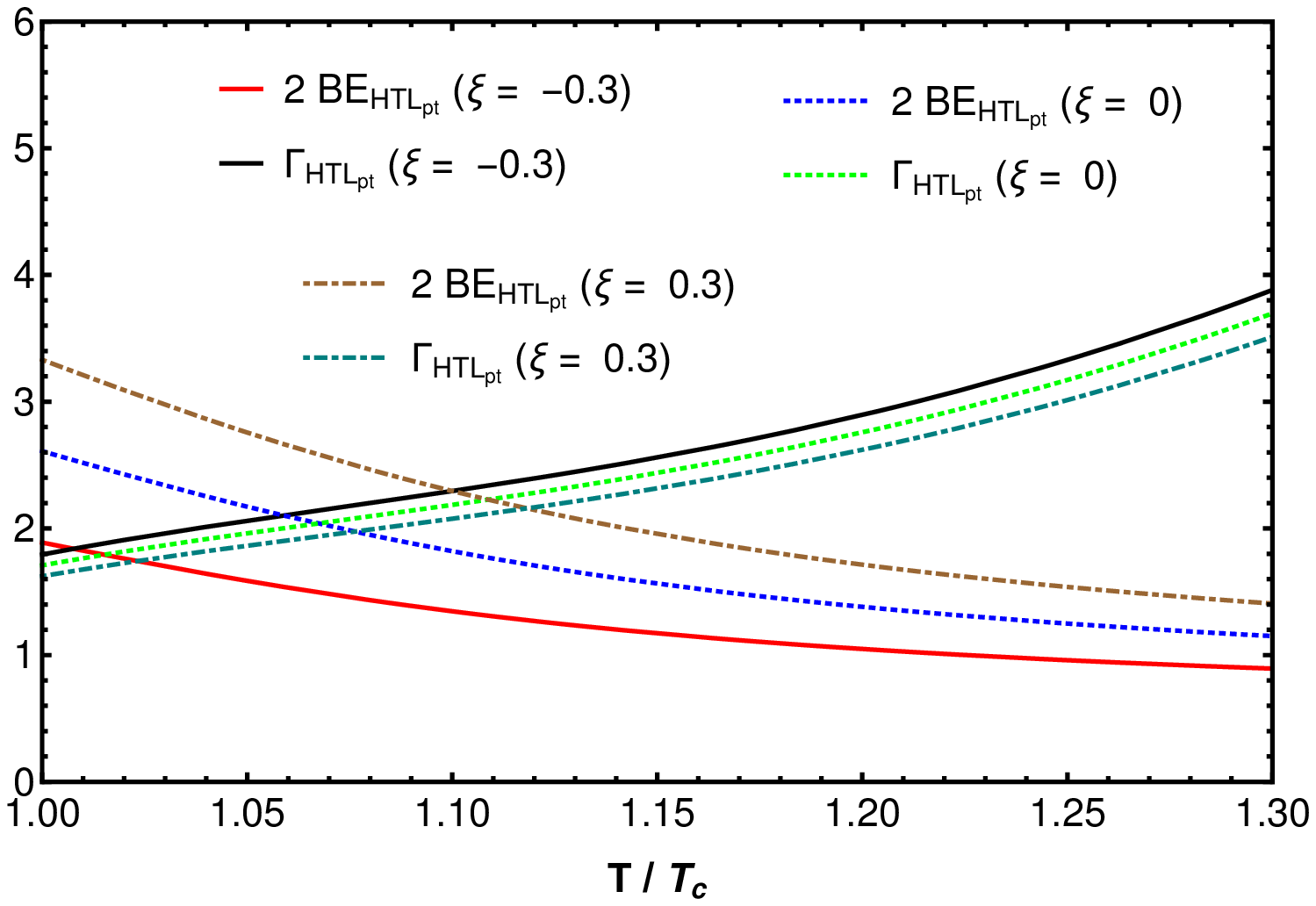}
    \hspace{-1mm}
    \includegraphics[height=5cm,width=5.8cm]{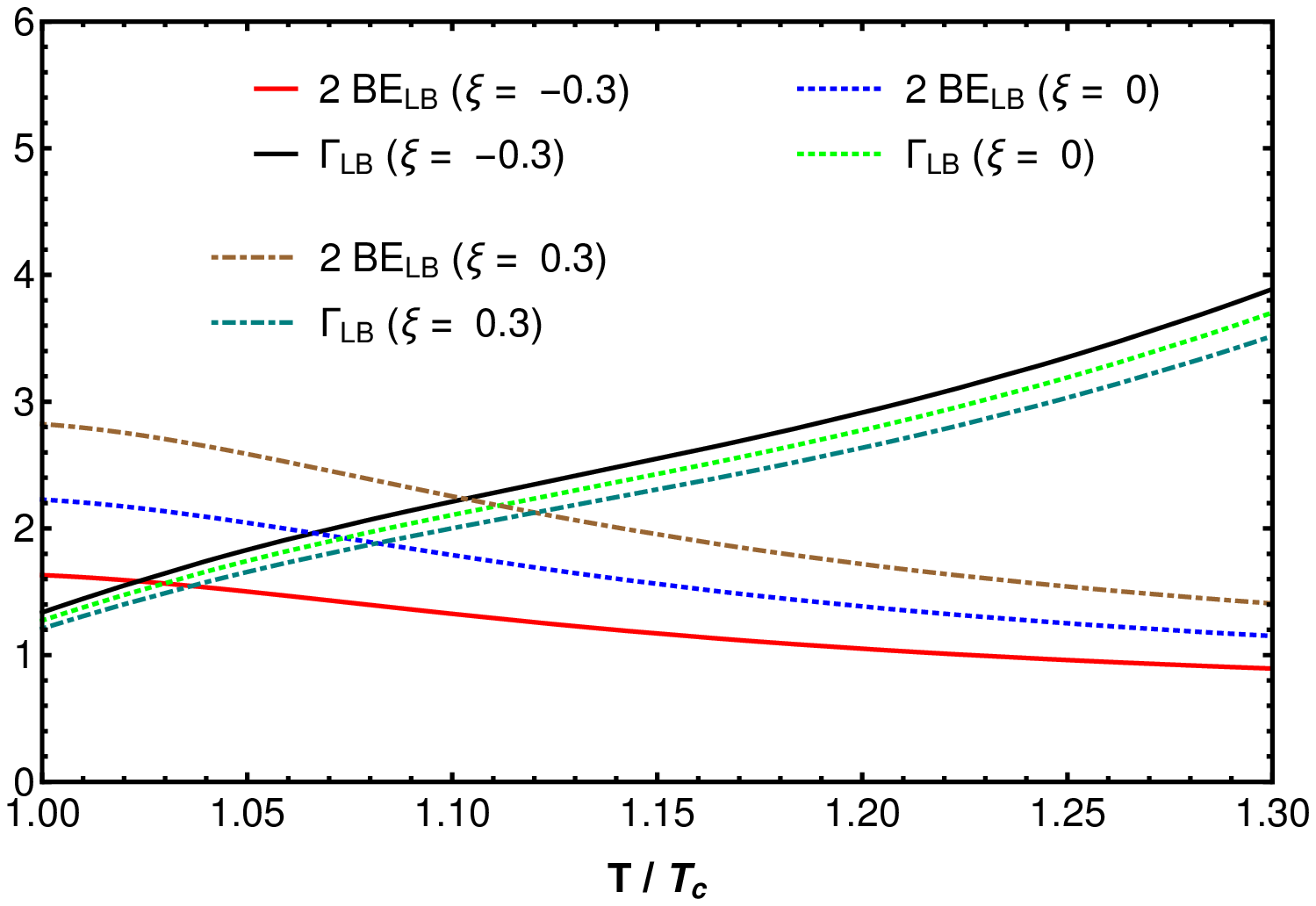}
     \caption{$\Gamma$, 2BE($E_b$) vs $T/T_c$ for $\Upsilon^{'}$ at $T_{c} = 0.17GeV$ with different $\xi$. We have plotted 
    the leading order (non-interacting) results (left panel) along with the 3-loop HTLpt (middle panel) and Lattice (right panel).}
    \label{fig:upsilon_p}
\end{figure*}

\subsection{Binding energy($E_b$) and thermal width ($\Gamma$)}

While considering the small anisotropy, one can solve the  Schr\"odinger equation and obtained the real part of the binding energy(BE or $E_b$)
by just considering the isotropic part with the first order perturbation in anisotropy parameter, $\xi$, as done in ~\cite{Margotta:2011ta, Strickland:2011aa, Thakur:2013nia}.
In this case, the real part of $E_b$ is realized as, 
 
\ba
{\text{Re}[E_b(T)]} &=&\bigg( \frac{m_Q~\sigma^2 }{m_{D}^4~ n^{2}} + \alpha~ m_{D} +\frac{\xi}{3}\Big(\frac{m_Q~\sigma^2 }{m_{D}^4 ~n^{2}}\nn
&+& \alpha~ m_{D}+ \frac{2~m_Q~\sigma^2 }{m_{D}^4~ n^{2}}\Big)\bigg).\ea

In the small-distance limit, the imaginary part of the potential can be considered as a perturbation to the vacuum potential ~\cite{Thakur:2013nia},
that provides an estimate for the thermal width for a particular resonance state given as,

\ba
\Gamma(T) = - \int d^3{\bf{r}}\, \left|\Psi({{r}})\right|^2{\rm{Im}}~V({\bf{r}}).
\label{Gamma}
\ea
 The medium potential, at the high temperature, has the long-range Coulombic tail that dominates over all the other terms.
Owing this fact, one can choose the $\Psi(r)$ as the Coulombic wave function. The Coulombic wave function for ground state ($1s$,
corresponding to $n=1$ ($J/\psi$ and $\Upsilon$)) and the first excited state ($2s$, corresponding to $n=2$ ($\psi'$ and
$\Upsilon'$)), respectively given as
\ba
\Psi_{1s}(r)&=& \frac{1}{\sqrt{\pi a_0^3}}e^\frac{-r}{a_0},~~\Psi_{2s}(r)= \frac{1}{4\sqrt{2\pi a_0^3}}\left(2-\frac{r}{a_0}\right)e^\frac{-r}{2a_0}.\nn
\label{psi}
\ea
where, $a_0$=$2/(\alpha ~m_Q)$ is the Bohr radius of the quarkonia system. Now using Eq.\ref{Gamma}, we have
 \ba
 \Gamma_{1s/2s}(T) &=&m_D^2 T\int d^3 {\bf r}|\Psi_{1s/2s}(r)|^2\bigg[\frac{\alpha}{3} \Big\{\frac{ \xi }{60} (7-9 \cos 2 \theta_r)-1\Big\}\nn
&+&\sigma~r^2 \Big\{\frac{\xi}{35}  \left(\frac{1}{9}-\frac{1}{4} \cos 2 \theta_r \right)-\frac{1}{30}\Big\}\bigg] r^2\log \left(\frac{1}{rm_D}\right).\nn
\ea
Rewriting the above equation as,

\ba
\Gamma_{1s}(T) =\left(\frac{\xi}{3} -2\right)m_D^2 T\bigg[{\alpha I_1}+\sigma I_2\bigg],
\label{Gamma1s}
\ea
where
\ba
I_{1}\ &=&\frac{12 \log \left(\frac{\alpha m_Q}{m_D}\right)+12 \gamma -25}{6 ~\alpha^2~  m_Q^2}
\label{I1}
\ea 
and 
\ba
I_{2}\ &=&\frac{3 \left(20 \log \left(\frac{\alpha m_Q}{m_D}\right)+20 \gamma -49\right)}{10~ \alpha^4~ m_Q^4}.
\label{I2}
\ea
Ultimately, the thermal width for $1s$-state appears as,
\ba
\Gamma_{1s}(T)&=&\frac{m_D^2~T~ (\xi -6) }{90~ \alpha ^4 ~m_Q^4} \bigg(60 \left(\alpha ^3 m_Q^2+3~ \sigma \right) \log \left(\frac{\alpha~ m_Q}{m_D}\right)\nn
&+&5 \left(12 \gamma -25\right) \alpha^3 m_Q^2+9 \left(20 \gamma -49\right) \sigma \bigg).
\label{Gamma1ss}
\ea
 It is important to note that in Ref.~\cite{Thakur:2013nia} while considering up to leading logarithmic order of imaginary potential,
the authors have taken the width also up to leading logarithmic. Thus, they consider the dissociation width of the following form for $1s$-state,
 \ba
 \Gamma_{1s}(T) &=&  T\bigg(\frac{4}{\alpha ~m_Q^2}+\frac{12\sigma}{\alpha^4~m_Q^4}\bigg) 
 \bigg(1-\frac{\xi}{6}\bigg)m_D^2 \log\left(\frac{m_D}{\alpha~ m_Q}\right).\nn
 \label{Gamma1sl}
 \ea 
 
In the present case, it has been observed that the additional terms, other than the leading logarithmic term, also contribute significantly.
Hence, we consider the full expression of the width given in Eq.~\ref{Gamma1ss}.  Note that, the authors in, Ref.~\cite{Thakur:2013nia} have
taken the normalization constant, $C_{\xi}$ equals to unity which is different in our case and has a remarkable contribution. Hence, we have
modified expression as one can see in Eq.\ref{Gamma1sl}, and other places as well. 
 
 For $2s$-state, we have
 \ba
\Gamma_{2s}(T) &=&\frac{T (\xi -6)}{45~ \alpha^2~m_Q^2}\bigg(35 (12 \gamma -31) \alpha + \frac{72 (160 \gamma -447) \sigma }{\alpha ^2~m_Q^2}\nn
&+&60 \left(7 \alpha +\frac{192~ \sigma }{\alpha ^2 m_Q^2}\right) \log \left(\frac{\alpha ~m_Q}{2~ m_D}\right)\bigg)m_D^2,
\label{Gamma2s}
\ea
 and the leading logarithmic order for $2s$-state is given as,
 \ba
 \Gamma_{2s}(T) =  \frac{8 ~m_D^2 T}{\alpha^4~ m_Q^4}\Big(1-\frac{\xi}{6}\Big)\bigg(7 \alpha ^3 m_Q^2+192 ~\sigma \bigg)\log \left(\frac{2~ m_D}{\alpha ~m_Q}\right).\nn
 \ea
 Again, in this case, we follow the solution given in Eq.~\ref{Gamma2s}, to calculate the binding energy.

Now, we have the real parts of binding energies, $E_b(T)$ (BE) as well as the thermal width, $\Gamma(T)$ for both the states.
Exploiting the criteria discussed earlier, we can plot twice the binding energy along with the thermal width and obtain the 
dissociation temperature as a point of their intersection. In the next section, we shall discuss the important results in details.
 
\section{Results and Discussion}
\label{RD}

In the present analysis, the various quantities have been obtained and the results are plotted, while considering the week anisotropy in the hot QCD plasma with the 
fixed critical temperature, $T_c = 0.17 ~$GeV. For prolate, $\xi=-0.3$ and for oblate, $\xi=0.3$ has been considered whereas for the isotropic one we have, $\xi=0$.
The EoSs that are employed here are symbolized as: the 3-loop HTL perturbative results are denoted with HTLpt while the lattice results are shown as LB 
and LO refers to the leading order (ideal case or the non-interacting case).

In Fig.~\ref{fig:realV}, the real part of the medium modified potential has been plotted with respect to $r$, using Eq.~\ref{eq:realv}, at temperature, $T = 3T_{c}~$GeV. 
In the LO case, the potential is seen to be less negative in contrast to non-ideal cases, at both parallel, $\theta_r = 0$ (left panel) and perpendicular,
$\theta_r = \pi/2$ (right panel).
For $\xi=0.3$, the numbers are slightly larger as compared to the $\xi=0$. The numbers for the prolate case, $\xi=-0.3$ are found to be the smallest among them.
 For $\theta_r =\pi/2$, as compared to $\theta_r =0$, the results are found to be similar but have slightly larger
separation for different anisotropies. This shows that the real part of the potential is marginally affected with the presence of anisotropy, as one traverse from the 
longitudinal plane to the transverse plane.  Also, we have obtained the Cornell potential from the modified one in the limit, $T\rightarrow 0$
with constant $\alpha$ and $\sigma$ and plotted in the same figures (Fig.~\ref{fig:realV}(a) and ~\ref{fig:realV}(b)) at $\alpha = 0.3$ and $\sigma = 0.184 ~GeV^2$.

In a similar way, the imaginary part of the medium modified potential, within the limit, $r m_D\ll1$ and considering the leading order in, $\xi$ has been plotted in
Fig.~\ref{fig:imagV}, with the same parameters as discussed in the case of real part, using Eq.~\ref{eq:imagv}. For the smaller values of $r$, the imaginary part of
the medium modified potential is found to be positive. As the effective radius increases, there occurs a crossover to the negative values. In both the cases, 
$\theta_r =\pi/2$ and $\theta_r =0$, the non-ideal EoSs are following the same pattern  as the ideal one. The effect of anisotropy is found to have a less impact on the 
 imaginary part of the potential as compared to the real one.

\begin {table}[H]
\caption {Ideal (non-interacting EoS) results for all three  prolate, isotropic and oblate cases} \label{tab:LO} 
\begin{center}
\begin{tabular}{ |p{2cm}||p{2cm}|p{2cm}|p{2cm}|  }
\hline
 \multicolumn{4}{|c|}{LO results} \\
 \hline
 \multicolumn{4}{|c|}{Temperatures are in the unit of $T_c$}\\
\hline
 \hline 
 States $\downarrow$ & $\xi = -0.3$& $\xi = 0.0$ &$\xi = 0.3$\\
 \hline
 $\Upsilon$   & 2.861  &2.964& 3.062\\
 $\Upsilon'$& 1.447  & 1.478   &1.508\\
 $J/\psi$ &1.487 &1.520& 1.551\\
 \hline
  \end{tabular}
 \end{center}
\end{table}

As discussed earlier, the dissociation temperature has been obtained by employing the criterion that says, the temperature at which twice the binding
energy (real part) equals the thermal width, causes dissociation of quarkonia, is the dissociation temperature.
In Fig.~\ref{fig:jpsi}, \ref{fig:upsilon} and \ref{fig:upsilon_p}, respectively, the thermal width of $J/\psi$,  $\Upsilon$ and $\Upsilon^{'}$ have been plotted,
along with twice the real part of their corresponding BE. In each case, LO is shown in the left, HTLpt in the middle and LB in the right panel. 
To plot them, the masses for $J/\psi,~\Upsilon~ \text{and}~\Upsilon^{'}$ are taken as $3.096~$GeV, $9.460~$GeV and $ 10.023~$GeV, respectively,
as calculated in ~\cite{Aulchenko:2003qq,Artamonov:1983vz,Barber:1983im}. In all the plots, for oblate cases, $\xi = 0.3$, the intersection points are found to be
the larger as compared to the isotropic cases, $\xi = 0$. The numbers for prolate cases, $\xi = -0.3$, observed to be the least among them. 
The results for the various EoSs are shown separately in the various Tables. 
The LO results for different anisotropies are presented in Table~\ref{tab:LO},  3-loop HTL perturbative calculation results are in Table ~\ref{tab:HTLpt}, 
and $(2+1)$- lattice results are shown in Table~\ref{tab:LB}. In the LO case, one can observe that the dissociation 
temperature is higher for $\Upsilon$ ($1s$- state), as compared to $J/\psi$ ($1s$- state), while the excited states, $\Upsilon^{'}$ ($2s$- state) has
lowest dissociation temperature. This hierarchy has been observed in all the three oblate, prolate and the isotropic cases.

\begin {table}[H]
\caption { HTL perturbative results for all three  prolate, isotropic and oblate cases} \label{tab:HTLpt} 
\begin{center}
\begin{tabular}{ |p{2cm}||p{2cm}|p{2cm}|p{2cm}|  }
\hline
 \multicolumn{4}{|c|}{3-loop HTLpt} \\
 \hline 
  \multicolumn{4}{|c|}{Temperatures are in the unit of $T_c$}\\
\hline
 \hline
 
 States $\downarrow$ & $\xi = -0.3$& $\xi = 0.0$ &$\xi = 0.3$\\
 \hline
 $\Upsilon$   & 2.427  &2.540& 2.639\\
 $\Upsilon'$& 1.008  & 1.067   &1.118\\
 $J/\psi$ &1.054 &1.119& 1.172\\
 \hline
 \end{tabular}
 \end{center}
\end{table}

A similar pattern is observed while taking the hot QCD medium effects into consideration, either through HTL perturbative results  or lattice 
simulation results as shown in, Table~\ref{tab:HTLpt} and Table~\ref{tab:LB}, respectively. But the essential point is that, as one gets
closer to the realistic picture by including the hot QCD medium interaction effects, a fall in the values of dissociation temperatures have been observed.
The non-ideal EoSs has almost overlapping numbers but to be more precise, the 3-loop HTLpt results are found to be the smallest among them.

\begin {table}[H]
\caption {Lattice simulation results for all three  prolate, isotropic and oblate cases} \label{tab:LB} 
\begin{center}
\begin{tabular}{ |p{2cm}||p{2cm}|p{2cm}|p{2cm}|  }
\hline
 \multicolumn{4}{|c|}{Lattice Bazabov(2014)} \\
 \hline
 \multicolumn{4}{|c|}{Temperatures are in the unit of $T_c$}\\
\hline
 \hline 
 States $\downarrow$ & $\xi = -0.3$& $\xi = 0.0$ &$\xi = 0.3$\\
 \hline
 $\Upsilon$   & 2.451  &2.564& 2.665\\
 $\Upsilon'$& 1.023  & 1.074   &1.120\\
 $J/\psi$ &1.063 &1.121& 1.172\\
 \hline
 \end{tabular}
 \end{center}
\end{table}

The numbers for dissociation temperatures are found to be consistent with those given in Ref.~\cite{Thakur:2013nia,Mocsy:2007jz}. Specifically, while implementing
the interacting EoSs, the numbers are observed to be closer. For each state, we displayed a contrast in, Table~\ref{tab:der} by calculating the decrease 
(in percentage ($\%$)) in the dissociation temperatures due to the presence of hot QCD medium effects. It is found that the dissociation temperatures while
incorporating hot QCD medium effects, has been lowered down by around $13\%$ to $31\%$. The excited state, $\Upsilon^{'}$ has reduce almost twice as compared to the
ground state $\Upsilon$.

\begin {table}[H]
\caption {Drop in the dissociation temperature from LO results while applying HTLpt and Lattice.} \label{tab:der}
\begin{center}
\begin{tabular}{ |p{2cm}||p{2cm}|p{2cm}|p{2cm}|  }
 \hline
 \multicolumn{4}{|c|}{Change in percentage ($\%$) }\\
\hline
 \multicolumn{4}{|c|}{Using HTLpt} \\
 \hline 
States$\downarrow$ & $\xi = -0.3$& $\xi = 0.0$ &$\xi = 0.3$\\
 \hline
 $\Upsilon$& 15.2  &14.3&13.8\\
 $\Upsilon'$& 30.3  & 27.8&25.9\\
 $J/\psi$&29.1&26.7&24.4\\
\hline
 \multicolumn{4}{|c|}{Using Lattice}\\
 \hline
 $\Upsilon$&14.3&13.5& 13.0\\
 $\Upsilon'$& 29.3&27.3&25.7\\
 $J/\psi$ &28.5&26.6& 24.4\\
 \hline
 \end{tabular}
 \end{center}
\end{table}

It is to note that, we also tried for $\psi^{'}$ ($2s$- state, whose mass is $3.686~$GeV), but did not find any intersection
point above the critical temperature ($T_c$). Based on the earlier discussion, this behaviour is expected because the excited
states decay at low temperature than their corresponding ground state ($J/\psi$, which is found to decay at very close to $T_c$).
A similar suppression patter has already been seen in the case of $\Upsilon^{'}$ and $\Upsilon$.

\section{Summary and Conclusion}
\label{Con}
The dissociation temperatures for the bottonium and the charmonium (ground ($1s$) as well as first excited ($2s$))
states have been obtained using the medium modified inter-quark potential in the anisotropic hot QCD medium.
The real/imaginary part of the heavy-quark potential is obtained in terms of real/imaginary part of the complex permittivity.
The real part of the medium modified potential causes a
dynamical screening of color charge that leads to the temperature dependent binding energy. Whereas the imaginary part
of the same leads to the temperature dependent thermal dissociation width.
 It has been observed that with the increase in temperature binding energy of the heavy quarkonia decreases while the thermal dissociation width increases.
Exploiting the criteria employed here, the dissociation temperature for each state has been calculated where twice the
binding energy equals the thermal dissociation width. 

The hot QCD medium interaction effects have been incorporated through the Debye mass by employing the EQPM. 
Considering the above fact, the Debye mass has been normalized in both the medium to make it intact from the effects
of the anisotropy so that the impact of various EoSs can be seen clearly through it. The results coming out after
incorporating the medium interaction effects are found to be smaller in magnitude as compared to the non-/weekly
interacting ideal one. It has been further noticed that the finite momentum-space
anisotropy in prolate, $\xi<0$ case decreases while in the oblate, $\xi>0$ case increases the dissociation temperature 
as compared to isotropic one, $\xi=0$ for all states taken into account. It has also been observed that in both the
cases (charmonium and bottonium) the excited states dissociate earlier (at low temperature) than their corresponding ground state. 
Furthermore, we have not found intersection point above, $T_c$ for the $\psi^{'}$-state.
Finally, we observed that both the anisotropy and the hot QCD medium effects present in EoS play a significant role in deciding the fate of heavy-quarkonia states 
 in the hot QCD/QGP medium.
 
To extend the present work, we aim to incorporate the viscous effects and study the dissociation of heavy quarkonia in
hydro-dynamically expanding viscous QGP. Apart from that the collisional effects on the quarkonia suppression will be carried out in the near future.

\section*{Acknowledgement}
We would like to acknowledge people of INDIA for their generous support for the research in fundamental sciences in the country.
I. Nilima acknowledges IIT Gandhinagar for the academic visit and hospitality during the course of this work.

\end{document}